\documentclass[aps,manuscript,showpacs,showkeys,prbbib]{revtex4}
\usepackage{amsmath}
\usepackage{amssymb}
\usepackage{graphicx}
\usepackage{dcolumn}
\usepackage{bm}

\begin{document}
\draft

\title{Escape Rates of the H$\mathrm{\acute{e}}$non-Heiles
System}

\author{H.J. Zhao and  M.L. Du} \email{
duml@itp.ac.cn}

\address{Institute of Theoretical Physics, Chinese Academy of Sciences,
P.O.Box 2735, Beijing 100080, China}

\date{\today}

\begin{abstract}
A particle in the H$\mathrm{\acute{e}}$non-Heiles potential can
escape when its energy is above the threshold value
$E_{th}=\frac{1}{6}$. We report a theoretical study on the the
escape rates near threshold. We derived an analytic formula for the
escape rate as a function of energy by exploring the property of
chaos. We also simulated the escaping process by following the
motions of a large number of particles. Two algorithms are employed
to solve the equations of motion. One is the Runge-Kutta-Fehlberg
method, and another is a recently proposed fourth order symplectic
method. Our simulations show the escape of
H$\mathrm{\acute{e}}$non-Heiles system follows exponential laws. We
extracted the escape rates from the time dependence of particle
numbers in the H$\mathrm{\acute{e}}$non-Heiles potential. The
extracted escape rates agree with the analytic result.

\par


\par

\end{abstract}

 \maketitle


\section{Introduction and model}
Bauer and Bertsch\cite{Bauer} studied the decay laws of chaotic and
non-chaotic billiards with windows. They found the number of
classical particles remaining inside chaotic billiards decreases
exponentially, but no-chaotic billiards decay according to power
laws. The results suggest that the exponential law is connected to
the chaotic dynamics. However, an exception is the circular
billiard, which is integrable but decays
exponentially\cite{legrand}. Experimental studies on the decay laws
of an elbow cavity using microwaves have also been
reported\cite{Doron}. We consider the
H$\mathrm{\acute{e}}$non-Heiles system\cite{Henon,Gutzwiller} with
the following Hamiltonian
\begin{eqnarray}
H &=& \frac{1}{2}(p_{x}^{2}+p_{y}^{2})+U(x,y), \\ \nonumber
 U(x,y)&=& \frac{1}{2}(x^{2}+y^{2})+x^{2}y-\frac{1}{3}y^{3},
\end{eqnarray}
where $x$ and $y$ are the coordinates, $p_{x}$ and $p_{y}$ are the
momenta. The mass of the particle is set to one for convenience.
This system exhibits both regular motion and chaotic motion
depending on the energy of the system, and it has been studied from
statistical, semiclassical and other
perspectives\cite{Gustavson,Berdichevsky,Noid,Delos,Zhigang}.
Recently Brack et al have calculated the density of states above
threshold\cite{Brack}, but the escape problem has not been addressed
so far.

Numerical studies show the motion of H$\mathrm{\acute{e}}$non-Heiles
system is regular for $E<1/12$. When $E$ is greater than 1/12, the
fraction of chaotic region in phase space increases with increasing
energy until $E=1/6$ the whole phase space is chaotic. $E_{th}=1/6$
is the threshold energy of this system. When $E \geq E_{th}$, a
particle in the potential well can escape. Fig. 1 shows the contours
of the potential $U(x,y)$. There are three saddle points
$P_{1}(x=0,y=1)$, $P_{2}(-\sqrt{3}/2,-1/2)$ and
$P_{3}(\sqrt{3}/2,-1/2)$. All contours with energy less than 1/6 are
closed. A particle with energy less than 1/6 always moves inside the
closed contour and it remains in the well. The contour with $E=1/6$
is the equilateral triangle $P_{1}P_{2}P_{3}$. The contours with
energies larger than 1/6 are not closed. There are three openings at
the three saddle points. A particle with energy above 1/6 can escape
from the well via the three openings. Because this system is chaotic
above threshold, the escape in the H$\mathrm{\acute{e}}$non-Heiles
system should also follow an exponential law. Assuming $N(0)$ random
particles with the same energy in the
H$\mathrm{\acute{e}}$non-Heiles well at $t=0$, the number of
particles at $t$ will be
\begin{equation}\label{2}
    N(t)=N(0)\exp(-\alpha t),
\end{equation}
where $\alpha$ is an energy dependent decay rate. The purpose of
this article is to verify Eq.(2) and to estimate $\alpha$ for
different energies.

\section{An escape rate formula}

We can derive a formula for the escape rate as a function of energy
above threshold by using chaotic property of the
H$\mathrm{\acute{e}}$non-Heiles system. We draw a line perpendicular
to the escape direction through each saddle point, they are line
$A_{1}B_{1}$,$A_{2}B_{2}$ and $A_{3}B_{3}$ in Fig. 1. For any energy
$E$ above threshold, we define the potential well of the
H$\mathrm{\acute{e}}$non-Heiles system to be the region restricted
by the three disconnected contour lines and the three straight lines
$A_{1}B_{1}$,$A_{2}B_{2}$ and $A_{3}B_{3}$. The motion in the well
is assumed to be ergodic because of chaos. The distribution on the
energy shell is given generally by
$\psi(q,p)=\frac{\delta(E-H(q,p))}{\int dqdp\delta
(E-H(q,p))}$,where $q,p$ are the coordinates and
momenta\cite{Gutzwiller}. For our two-dimensional system, it is easy
to work out the results: the distribution in (x,y) is uniform inside
the well, and once the particle's position is given, the magnitude
of the momentum is fixed by the Hamiltonian in Eq.(1) and the
direction of the momentum is uniformly distributed in $[0,2\pi]$. We
define the energy above threshold $\Delta E= E-E_{th}$. We use
$\theta$ to represent the direction of momentum relative to the y
axis. We use $S(\Delta E)$ to denote the area of the well. Then the
distribution in the variables $(x,y,\theta)$ can be expressed as
$\rho (x,y,\theta)=\frac{1}{2\pi S(\Delta E)}$. Given $N$ particles
in the well, the number of particles leaving the well through the
opening at the saddle point $P_{1}$ in unit time can be written as
$N\int dx \int_{-\pi/2}^{\pi/2}d\theta
\rho(x,y,\theta)|\mathbf{v}(x,y)|cos(\theta)$, where the integral in
x is along the line $A_{1}B_{1}$ and is restricted to the classical
allowed part. We note the three openings of the system are
symmetric. Therefore the number of particles leaving the well in
unit time from three openings are just three times of the above
result. The change of $N$ with respect to $t$ is
 \begin{eqnarray}
\frac{dN(t)}{dt} & = & -3N(t)\rho \int_{-\pi/2}^{\pi/2}cos(\theta)d\theta
\int_{-\sqrt{2\Delta E/3}}^{\sqrt{2\Delta E/3}}\sqrt{2(\Delta E-3x^{2}/2)}dx\\
 & = & -2\pi\sqrt{3}\Delta E \rho N(t),
\end{eqnarray}
which gives the escape rate $\alpha(\Delta E)=\frac{\sqrt{3}\Delta
E}{S(\Delta E)}$.

There is no analytical formula for the area of the well $S(\Delta
E)$. We have applied Monte Carlo method to calculate the area as a
function of $\Delta E$. The numerical results are represented by
dots in Fig.2. We found the numerical results can be represented
vary well by the quadratic polynomial
\begin{equation}\label{4}
    S(\Delta E)=S_{0}+S_{1}\Delta E +S_{2}(\Delta E)^2
\end{equation}
 where $S_{0}=\frac{3\sqrt{3}}{4}$ is the area of equilateral triangle $P_{1}P_{2}P_{3}$.
By fitting Eq.(5) to the numerical results in Fig.2 using least
squares, we have determined the values for the other two
coefficients, $S_{1}=9.656$ and $S_{2}=-22.61$. The line in Fig.2 is
the fitted quadratic polynomial. We finally have the formula for the
escape rate in the H$\mathrm{\acute{e}}$non-Heiles system,

\begin{equation}\label{5}
    \alpha(\Delta E)= \frac{\sqrt{3}
    \Delta E}{S_{0}+S_{1}\Delta E +S_{2}(\Delta E)^{2}}.
\end{equation}

Very close to the threshold, a power expansion of Eq.(6) may be
useful. Define
\begin{equation}\label{6}
    \alpha(\Delta E)=\sum_{1}^{\infty}B_i{\Delta E}^i.
\end{equation}
The coefficients $B_i$ can be expressed in $S_0,S_1$ and $S_2$. They
are $B_{1}=\frac{\sqrt{3}}{S_{0}}$,
$B_{2}=-\frac{\sqrt{3}S_{1}}{S_{0}^{2}}$, and others can be obtained
from the following iteration formula,
$B_{i}=-\frac{S_{1}B_{i-1}+S_{2}B_{i-2}}{S_0},i=3,4,...$. The
numerical values for the first four coefficients in the power
expansion of Eq.(6) are $B_1=4/3,B_2=-9.9115,B_3=96.8743$, and
$B_4=-892.547$. The first term in the power expansion is
proportional to $\Delta E$, it gives the scaling of the escape rate
in the H$\mathrm{\acute{e}}$non-Heiles system just above threshold.

\section{Numerical simulations}

In our numerical simulations of the escape processes, we follow a
large number of particles. We monitor the number of particles $N(t)$
remaining in the potential well as a function of time and then
extract the escape rate. For any energy above threshold, we
initially place $N(0)$ particles in the well according to the
distribution $\rho (x,y,\theta)=\frac{1}{2\pi S(\Delta E)}$. This
distribution sets the initial conditions for the particles. The
trajectory of each particle is then followed by numerically solving
the Hamilton's equations. We used two algorithms to integrate
Hamilton's equation. Runge-Kutta-Fehlberg (RKF) is the first
algorithm\cite{RKF}. In this algorithm the error in each step can be
controlled by setting the relative tolerance and the absolute
tolerance. In all our calculations, we set the absolute tolerance to
$10^{-9}$. The second algorithm (CC) was proposed recently by Chin
and Chen\cite{SFI}, it is a fourth order forward symplectic
algorithm. It is generally believed that symplectic algorithms are
better and can follow the true dynamics longer because they preserve
the symplectic structures of Hamilton's equation. The explicit
algorithm for advancing the system forward from $t$ to $t+\epsilon$
is
\begin{eqnarray}
 \mathbf{p}_{1}&=&\mathbf{p}(i)+\frac{1}{6}\epsilon
\mathbf{F}(\mathbf{q}(i)) \\ \nonumber
  \mathbf{q}_{1}&=&\mathbf{q}(i)+\frac{1}{2}\epsilon \mathbf{p}_{1}\\\nonumber
  \mathbf{p}_{2}&=&\mathbf{p}_{1}+\frac{4}{6}\epsilon
\mathbf{\tilde{F}}(\mathbf{q}_{1}) \\\nonumber
  \mathbf{q}(i+1)&=&\mathbf{q}_{1}+\frac{1}{2}\epsilon \mathbf{p}_{2}\\\nonumber
\mathbf{p}(i+1)&=&\mathbf{p}_{2}+\frac{1}{6}\epsilon
\mathbf{F}(\mathbf{q}(i+1)).
\end{eqnarray}
Note $\textbf{F}=-\nabla U$ and
$\mathbf{\tilde{F}}=\textbf{F}+\frac{1}{48}\epsilon^{2}\nabla(|\textbf{F}|^{2})$
includes an correction to the original force. The time step size
$\epsilon$ can be varied to control integration errors.

In Fig. 3 we show the two trajectories calculated using the two
algorithms. The two trajectories start from the same initial
condition: the position is at ($x=0$, $y=0.16$), the energy is
$E=0.18$, and the direction of momentum is in the positive x-axis.
Fig.3(a) is the trajectory obtained using RKF algorithm with
relative tolerance $10^{-8}$. Fig.3(b) is the trajectory obtained
using CC method with a time step size $\epsilon=0.04$. We have
verified the accuracy of both calculations. When the time step size
and relative tolerance were reduced further, the trajectories did
not show noticeable change. Fig.3(a) and Fig.3(b) show clearly the
two trajectories obtained using the two algorithms stay close for
some time and then separate. In Fig. 3(a), the particle escapes at
$t=299$, while in Fig.3(b), the particle escapes at $t=172$. The
increased separation of the two trajectories calculated with two
different algorithms starting with the same initial condition
reflects the difficulty to follow chaotic motions for a long time.
Nevertheless we can still extract accurate escape rates as shown
below.

A large number of particles are used in the simulations of escape
process.  For example, we initially put $N(0)=15326$ particles with
the energy $\Delta E=0.0234$ above threshold according to $\rho
(x,y,\theta)=\frac{1}{2\pi S(\Delta E)}$ in the well. We advanced
this system by following the trajectory of each particle using RKF
method with relative tolerance $10^{-8}$. We recorded the number of
particles $N(t_i)$ remaining in the well in time step $\Delta
t=0.628$. Fig. 4(a) shows $N(t)$ as a function of time in log-linear
scale. Fig. 4(b),Fig. 4(c) and Fig. 4(d) show similar results for
different energies. One can see the curves in Fig. 4 in all the
cases are almost straight lines, indicating the escape of
H$\mathrm{\acute{e}}$non-Heiles system follows exponential laws. For
each energy, we can extract the escape rate from the simulated
$N(t)$. We used the simulated $N(t)$ from time $t=0$ to a time when
ten percent of the particles have escaped and fitted it to $ln
N(t)=c-\alpha t$ using least squares. The fitted parameter $\alpha$
is the extracted escape rate at the corresponding energy.

We compare in Fig.5 the extracted escape rates and the analytic
result in Eq.(6) as a function of energy above threshold. The solid
line is from the formula in Eq.(6). The circles are the numerical
results from RKF method with a relative tolerance $10^{-8}$, the
diamonds are the numerical results obtained from CC method (Ref.13)
with time step 0.04. We verified the numerical accuracy with smaller
tolerance and smaller step size, and we found the results did not
change. The dashed line also shows the result of the power expansion
of Eq.(6) with only the first four terms. The extracted results from
the two algorithms agree well. They also agree with the formula in
Eq.(6). The first four terms power expansion is accurate close to
threshold, when $\Delta E$ is greater than 0.05, it starts to
deviate from the analytic result in Eq(6) and from the numerical
results.

\section{Conclusions}

We have derived a formula in Eq.(6) for the escape rate of the
H$\mathrm{\acute{e}}$non-Heiles system. We also simulated the escape
process by following a large number of trajectories. We used a
symplectic algorithm and a non-symplectic algorithm to advance each
particle's trajectory in time. We monitored the number of particles
remaining in the potential well in time. We found the escape follow
exponential laws similar to the chaotic billiard
systems\cite{Bauer}. The extracted escape rates using the two
algorithms agree with each other, and and they also agree with the
analytic rate formula for H$\mathrm{\acute{e}}$non-Heiles system in
Eq.(6).

\begin{center}
{\bf ACKNOWLEDGMENTS}
\end{center}
\vskip8pt This work was supported by NSFC grant No. 90403028.


\clearpage

\begin{figure}
\includegraphics[scale=.40,angle=-0]{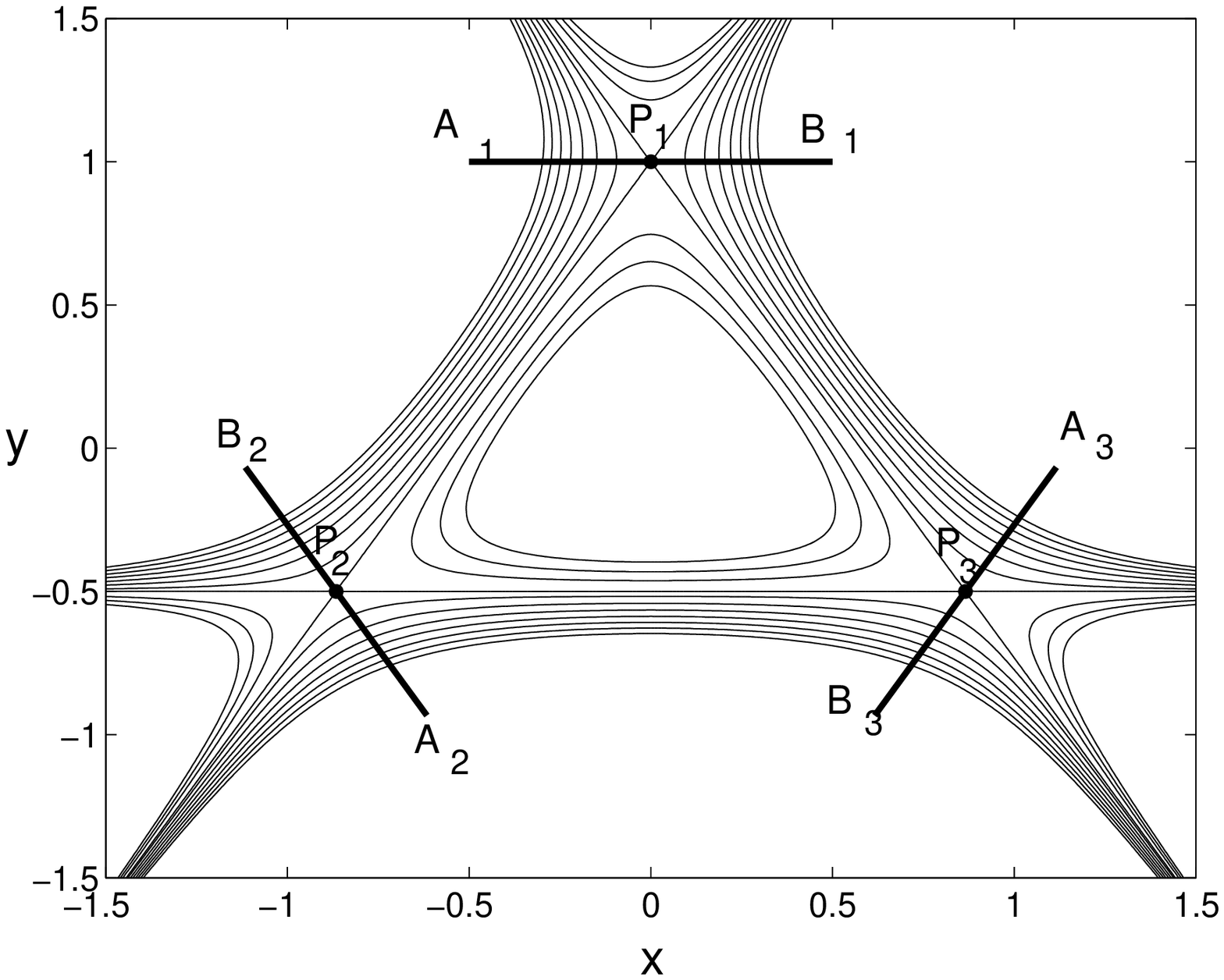}
\caption[]{ Equipotential lines of the function U(x,y) in Eq.(1).
The points $P_{1}$,$P_{2}$ and $P_{3}$ are saddle points.
 } \label{B2}
\end{figure}

\begin{figure}
\includegraphics[scale=.40,angle=-0]{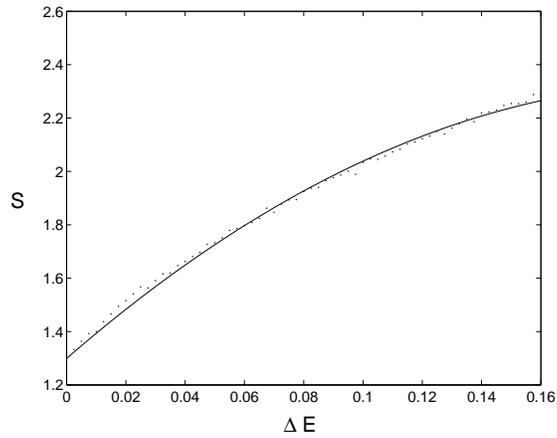}
\caption[]{The area of the well (region bounded by the three
disconnected contour lines and the three straight lines
$A_{1}B_{1}$,$A_{2}B_{2}$ and $A_{3}B_{3}$) as a function of energy
above threshold $\Delta E$. The dots are results calculated using
Monte Carlo method. The solid line is the fitted quadratic
polynomial in Eq.(5).
 } \label{B2}
\end{figure}

\begin{figure}
\includegraphics[scale=.80,angle=-0]{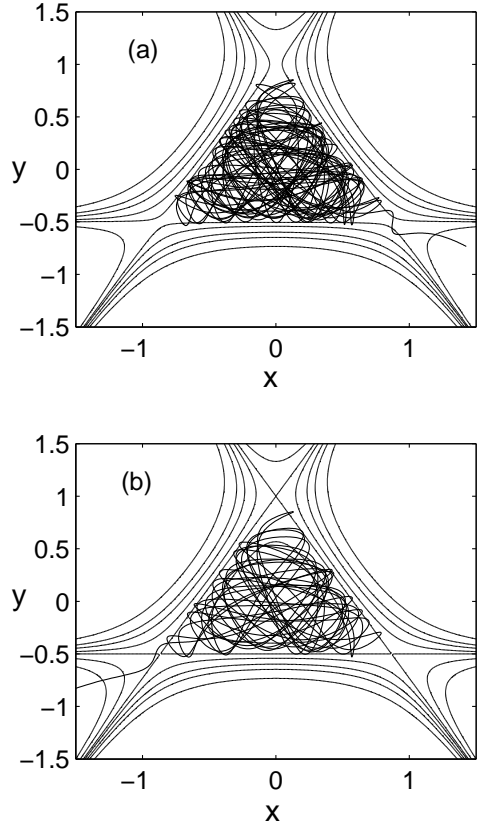}
\caption[]{Trajectories calculated from the two algorithms in the
H$\mathrm{\acute{e}}$non-Heiles system. Both trajectories start from
the position ($x=0,y=0.16$) with energy $E=0.18$, the initial
direction of momentum points to positive x-axis. (a) RKF method
(Ref.12) with relative tolerance $10^{-8}$ and time range (0-299).
(b) CC method of Chin and Chen (Ref.13) with step 0.04 and time
range (0-172).
    } \label{B2}
\end{figure}

\begin{figure}
\includegraphics[scale=.40,angle=-0]{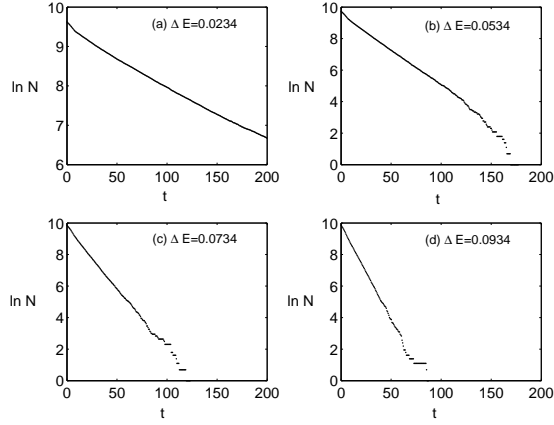}
\caption[]{The number of particles in the well as a function of time
for four energies above threshold. (a) $\Delta E=0.0234$,
$N(t=0)=15326$;(b)$\Delta E=0.0534$, $N(t=0)=17392$; (c)$\Delta
E=0.0734$, $N(t=0)=18885$;(d)$\Delta E=0.0934$, $N(t=0)=19942$. The
dots are numerical results obtained using RKF method with a relative
tolerance $10^{-8}$.
 } \label{B2}
\end{figure}

\begin{figure}
\includegraphics[scale=.40,angle=-0]{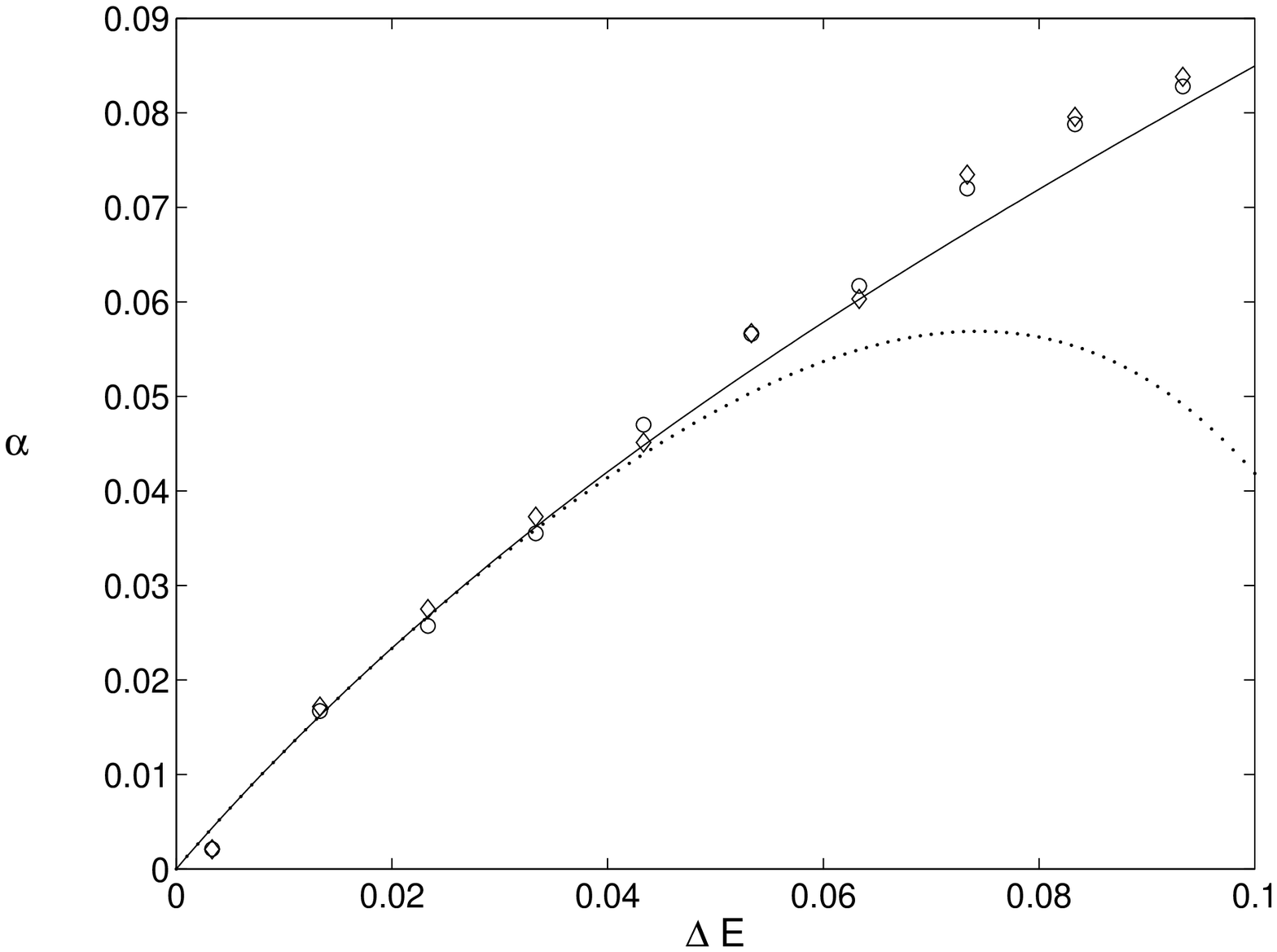}
\caption[]{The escape rate $\alpha$ vs $\Delta E$. The solid line is
the formula in Eq.(6). The circles are the numerical results from
RKF method with a relative tolerance $10^{-8}$, and the diamonds are
the numerical results obtained from CC method with a time step 0.04.
The dotted line is the power expansion of Eq.(6) with the first four
terms.
 } \label{B2}
\end{figure}



\begin{references}

\bibitem{Bauer}
W. Bauer and G. F. Bertsch, Phys. Rev. Lett.
 {\bf 65}, 2213 (1990).

\bibitem{legrand}
O. legrand and D. Sornette, Phys. Rev. Lett.
 {\bf 66}, 2172 (1991).

\bibitem{Doron}
E. Doron, U. Smilansky and A. Frenkel, Phys. Rev. Lett.
 {\bf 65}, 3072 (1990).


\bibitem{Henon}
M. H$\mathrm{\acute{e}}$non and C. Heiles, Astro. J.
 {\bf 69}, 73 (1964).

\bibitem{Gutzwiller}
M.C.Gutzwiller, Chaos in Classical and Quantum Mechanics (Springer,
New York, 1990).

\bibitem{Gustavson}
F. G. Gustavson, Astro. J.
 {\bf 71} 670 (1966).

\bibitem{Berdichevsky}
V. L. Berdichevsky and M. V. Alberti, Phys. Rev. A {\bf 44} 858
(1991).


\bibitem{Noid}
D. W. Noid and R. A. Marcus, J. Chem. Phys. {\bf 67} 559 (1977).

\bibitem{Delos}
J.B.Delos and R.T.Swimm, J. Chem. Phys. {\bf 47} 76 (1977).

\bibitem{Zhigang}
Zhigang Zheng, Gang. Hu and Juyuan Zhang,
 Phys. Rev. {\bf E52} 3440 (1995).

\bibitem{Brack}
M. Brack, J. Kaidel, P. Winkler, S. N. Fedotkin, arXiv/nlin/0511005

\bibitem{RKF}
M. C. W. H. Press, S. A. Teukolsky W. T. Vetterling and B. P.
Flannery, Numerical recipes in c: the art of scientific computing
(Cambridge University Press, Cambridge ,1988).

\bibitem{SFI}
S. A. Chin and C. R. Chen , arXiv/astro-ph/0304223.








\end{references}
\end{document}